\documentclass[sigconf,natbib=true,anonymous=false]{acmart}

\AtBeginDocument{%
  \providecommand\BibTeX{{%
    \normalfont B\kern-0.5em{\scshape i\kern-0.25em b}\kern-0.8em\TeX}}}

\copyrightyear{2024}
\acmYear{2024}
\setcopyright{acmlicensed}\acmConference[SIGIR '24]{Proceedings of the 47th International ACM SIGIR Conference on Research and Development in Information Retrieval}{July 14--18, 2024}{Washington, DC, USA}
\acmBooktitle{Proceedings of the 47th International ACM SIGIR Conference on Research and Development in Information Retrieval (SIGIR '24), July 14--18, 2024, Washington, DC, USA}
\acmDOI{10.1145/3626772.3657939}
\acmISBN{979-8-4007-0431-4/24/07}

\usepackage{paralist}
\usepackage{setspace}
\usepackage{multirow}

\begin{document}

\title{Neural Click Models for Recommender Systems}


\author{Mikhail Shirokikh}
\email{mscirokikh@yandex.ru}
\orcid{0009-0008-1771-0769}
\affiliation{%
  \institution{St. Petersburg University}
  \streetaddress{29 Line 14th (Vasilyevsky Island)}
  \city{St. Petersburg}
  \country{Russia}
  \postcode{199178}
}

\author{Ilya Shenbin}
\email{ilya.shenbin@gmail.com}
\orcid{0000-0002-6778-225X}
\affiliation{%
  \institution{PDMI RAS}
  \streetaddress{27 Fontanka}
  \city{St. Petersburg}
  \country{Russia}
  \postcode{191023}
}

\author{Anton Alekseev}
\orcid{0000-0001-6456-3329}
\affiliation{%
  \institution{PDMI RAS}
  \institution{St. Petersburg University}
  \city{St. Petersburg}
  \country{Russia}
}

\author{Anna Volodkevich}
\orcid{0009-0002-7958-0097}

\author{Alexey Vasilev}
\orcid{0009-0007-1415-2004}
\affiliation{%
  \institution{Sber AI Lab}
  \streetaddress{19 Vavilova Street}
  \city{Moscow}
  \country{Russia}
  \postcode{117997}
}

\author{Andrey V. Savchenko}
\orcid{0000-0001-6196-0564}
\affiliation{%
  \institution{Sber AI Lab}
  \institution{ISP RAS Research Center for Trusted Artificial Intelligence}
  \city{Moscow}
  \country{Russia}
}

\author{Sergey Nikolenko}
\orcid{0000-0001-7787-2251}
\affiliation{%
    \institution{PDMI RAS}
  \institution{St. Petersburg University}
  \institution{ISP RAS Research Center for Trusted Artificial Intelligence}
    \city{St. Petersburg}
    \country{Russia}
}

\renewcommand{\shortauthors}{Mikhail Shirokikh et al.}

\begin{abstract}
We develop and evaluate neural architectures to model the user behavior in recommender systems (RS) inspired by click models for Web search but going beyond standard click models. Proposed architectures include recurrent networks, Transformer-based models that alleviate the quadratic complexity of self-attention, adversarial and hierarchical architectures. Our models outperform baselines on the ContentWise and RL4RS datasets and can be used in RS simulators to model user response for RS evaluation and pretraining.
\end{abstract}



\begin{teaserfigure}\centering
  \includegraphics[width=.9\textwidth]{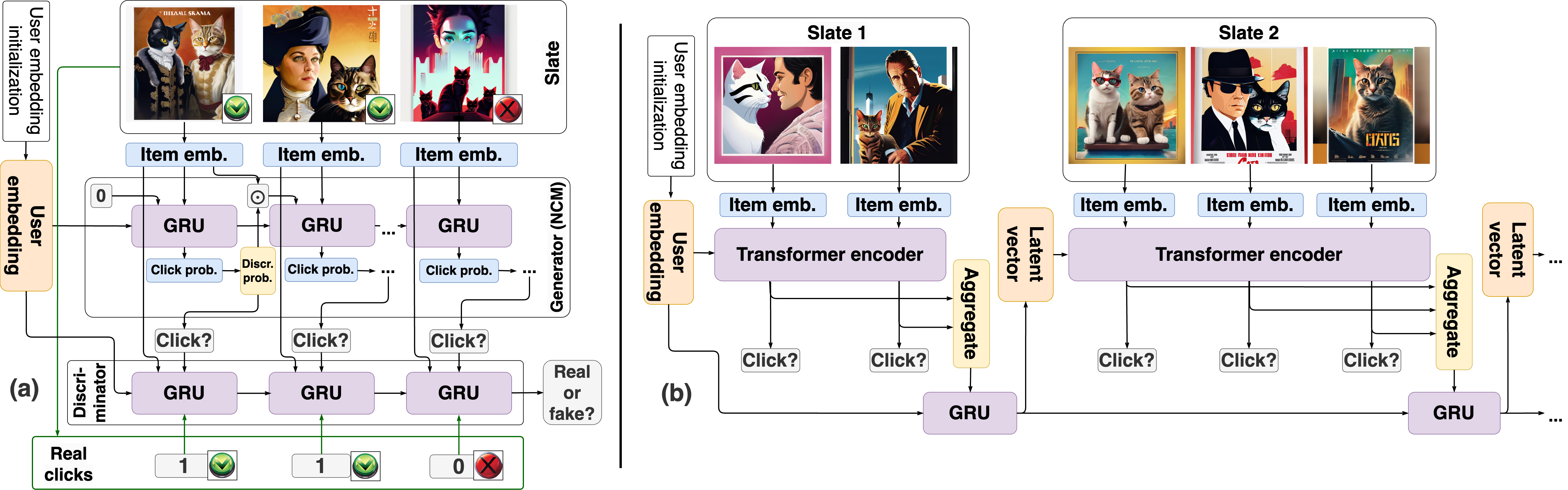}
  \caption{Two of the proposed models: (a) Adversarial Neural Click Model (AdvNCM); (b) two-stage Transformer + GRU model.}
  \Description{Two of the new user response models that we propose in this work: AdvNCM and Transformer+GRU.}
  \label{fig:teaser}
\end{teaserfigure}


\maketitle

\section{Introduction}

Recommender systems (RS) are an integral part of 
e-commerce, social networks, and content streaming platforms. They analyze user behavior and preferences to provide personalized recommendations, enhancing user experience and improving user engagement. However, evaluating their effectiveness 
and understanding their behavior in a real-world setting can be challenging. This is where \emph{simulators} for RS come into play, both for creating synthetic data for training and for modeling user behavior under recommendations produced by the RS,
allowing researchers and developers to study RS behavior and performance in controlled conditions. 
An important part of such simulators is the \emph{user response function} that generates synthetic user responses to recommended items.
In particular, a user response simulator has to be able to represent the structure of real-life online RS that can present recommended items in \emph{slates}, and a user \emph{session} may consist of several slates. 

Prior art on user response modeling can be broadly classified into:
\begin{inparaenum}[(i)]
    \item \emph{probabilistic models}, which were historically first~\cite{agrawal1994fast,deshpande2004item,Houkjr2006SimpleAR}, based on clustering~\cite{10.5555/2074022.2074076,10.1007/3-540-34416-0_29,Tso06,monti2019all}, learning skewed distributions~\cite{traupman2004collaborative}, resampling real data~\cite{marlin2005unsupervised}, context-aware models~\cite{DataGenCARS,doi:10.1155/2015/489264,10.1145/3151848.3151856,RodrguezHernndez2017TowardsTR}, causal models~\cite{lyu2022semi,10.1145/3404835.3462875} or inhomogeneous Poisson processes~\cite{mcinerney2021accordion};
    \item \emph{generative adversarial networks} (GAN), namely their variations for discrete and tabular data~\cite{10.1145/3474838,10.14778/3231751.3231757,10.5555/3454287.3454946,10.5555/3454287.3454946}, used to generate data for recommender systems in the form of user preference vectors~\cite{10.1145/3269206.3271743,10.1145/3308558.3313413}, user profiles~\cite{10.1007/978-3-030-16145-3_33}, individual interactions~\cite{10.1145/3292500.3330873,10.1016/j.eswa.2018.10.024,9240960}, temporal sequences~\cite{ZhaoPlastic18,10.1145/3240323.3240383}, or all of the above~\cite{bobadilla2023creating};
    \item full-scale \emph{synthetic data generators}, usually intended to serve as an environment for training a reinforcement learning agent, which is a popular framing for modern RS~\cite{CHEN2023110335,DBLP:journals/corr/abs-2109-10665,DBLP:journals/corr/abs-2101-06286,10.5555/3367243.3367400,gauci2018horizon,DBLP:journals/corr/abs-1812-02353}; here we note \emph{RecSim}~\cite{DBLP:journals/corr/abs-1909-04847}, \emph{RecSim NG}~\cite{DBLP:journals/corr/abs-2103-08057}, \emph{RecoGym}~\cite{DBLP:journals/corr/abs-1808-00720,JG20}, SOFA~\cite{huang2020keeping}, \emph{Virtual Taobao}~\cite{shi2019virtual}, and SARDINE~\cite{deffayet2023sardine} among other works~\cite{DBLP:journals/corr/abs-2101-04526}.
\end{inparaenum}

Our approach 
is based on \emph{click models} that 
study interactions between a search engine and a user
for user behavior simulation, web search evaluation, or improving ranking~\cite{chuklin2022click}; they assume that a user browses a page of results
and performs actions such as examining a snippet or clicking on a link. 
\emph{Cascade models} assume that a user views the page from top to bottom until they click; models such as the \emph{Dependent Click Model}~\cite{guo2009efficient} and \emph{Dynamic Bayesian Network}~\cite{chapelle2009dynamic} can handle multiple clicks per result page and use click data to extract document relevance cleaned of position bias.
Neural click models (NCM)~\cite{borisov2016neural} 
usually rely on recurrent neural networks (RNNs). NCMs are often more expressive and perform better but lack interpretability; 
we note
an NCM with arbitrary click order~\cite{borisov2018click}, session-level NCM~\cite{chen2020context}, graph-based click models~\cite{lin2021graph}, and NCM trained with imitation learning~\cite{dai2021adversarial}.
%
To the best of our knowledge, NCMs have not been applied as user response models for RS; in this work, we consider exploring the search results to be similar to exploring the list of recommended items (a slate).

We create and study a set of models to simulate user responses for slate-based recommendations, where responses to recommended items depend on previous interaction history and/or items shown in the current slate. 
We propose neural architectures to model the user response function, including several RNN- and Transformer-based architectures
and a hierarchical architecture that combines Transformers and RNN. 
Inspired by click models, we note that slate-based recommendation is similar to search results ranking, adapt some click models for our setting, and go beyond original click models with adversarial and Transformer-based architectures.
We present the results of a comprehensive experimental evaluation on two datasets, \emph{ContentWise Impressions}~\cite{contentwise} and RL4RS~\cite{rl4rsdataset}, that
contain information on impressions as well as positive interactions and organize recommended items in slates and sessions.
Our results indicate that the developed architectures are superior to the baselines and can serve as the basis for new RS synthetic data generators. 
The source code for our models is publicly available\footnote{\url{https://github.com/arabel1a/Neural-Click-Models-for-Recommender-Systems}}. 

\section{Datasets and baselines}\label{sec:data}

For evaluation, we need datasets that use slate-based recommendations and log impressions as well as click/purchase events; these requirements have restricted us to 
two public datasets.
\emph{ContentWise Impressions}~\cite{contentwise} contains logged recommendations and user interactions. It was collected from a video streaming service so most items (e.g., TV episodes) are grouped into \emph{series}. Fig.~\ref{fig:db}a shows the user screen layout, with more relevant recommendations (impressions) on top and to the left; user response is known for each pool of recommended items.
The \emph{Reinforcement Learning for Recommender Systems} dataset~\cite{rl4rsdataset} (RL4RS) 
comes from a mobile game by \textit{NetEase}, so impressions are shown with special unlock rules:
in the RL4RS-Slate dataset (Fig.~\ref{fig:db}b), new bundles of recommended items on a single page are shown after an unlock condition is fulfilled (the current list is sold out), while the RL4RS-SeqSlate dataset 
adds links between pages and asks to maximize the total reward over multiple pages.
RL4RS contains $140$K users and $283$ items with rich information: an anonymized user profile, user click history embeddings, item features with price, category, and more. 

\def\figonehei{.2\linewidth}
\def\figtwohei{.16\linewidth}
\def\figthreehei{.18\linewidth}

\begin{figure*}[!t]\centering
\setlength{\tabcolsep}{16pt}
\begin{tabular}{c|c|c}
\includegraphics[height=\figonehei]{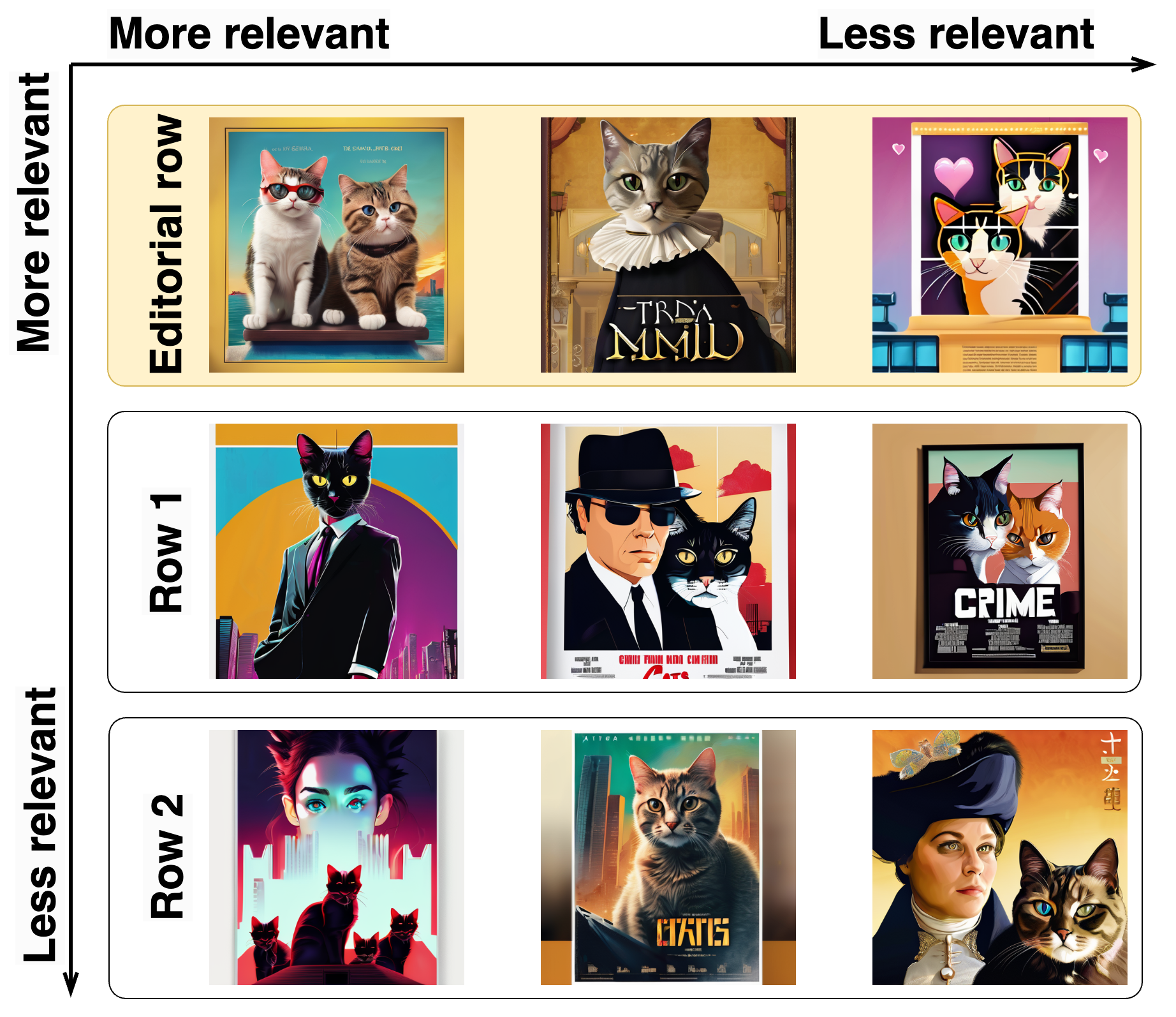} 
&
\includegraphics[height=\figonehei]{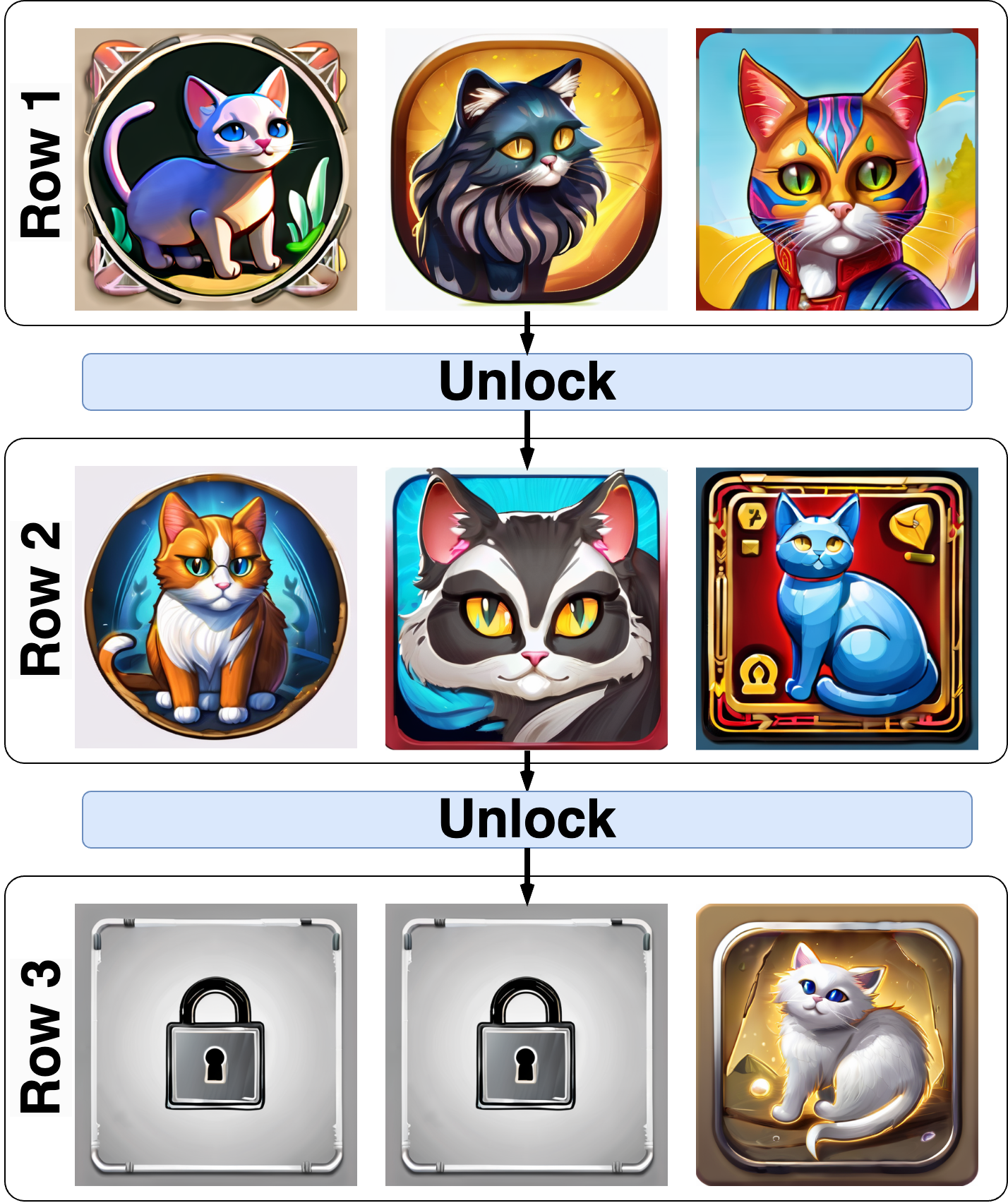} 
&
\includegraphics[height=\figonehei]{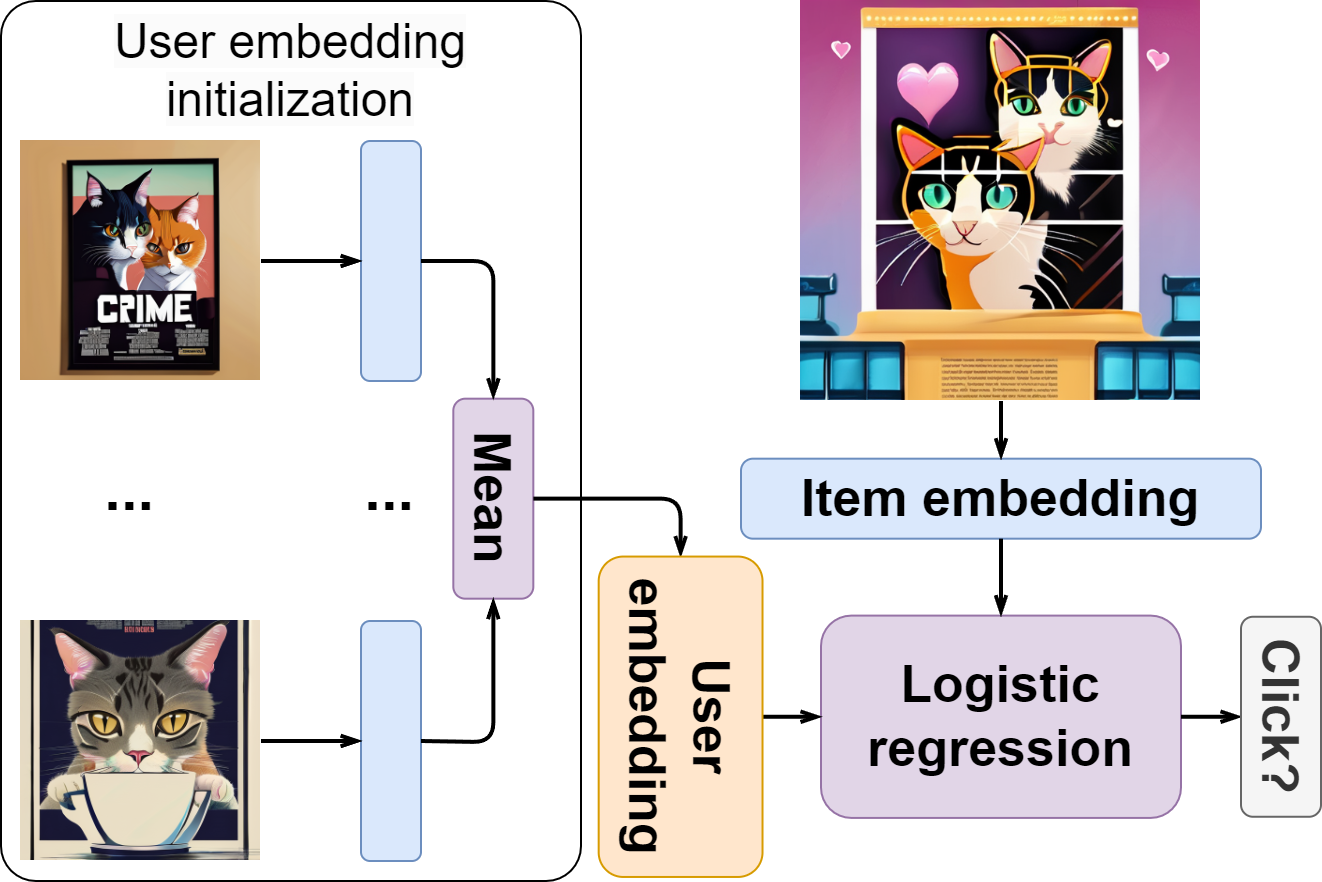} 
\\ (a) ContentWise sample slate & (b) RL4RS sample slate & (c) Logistic regression baseline
\end{tabular}\vspace{.3cm}

\begin{tabular}{c|c}
\includegraphics[height=\figtwohei]{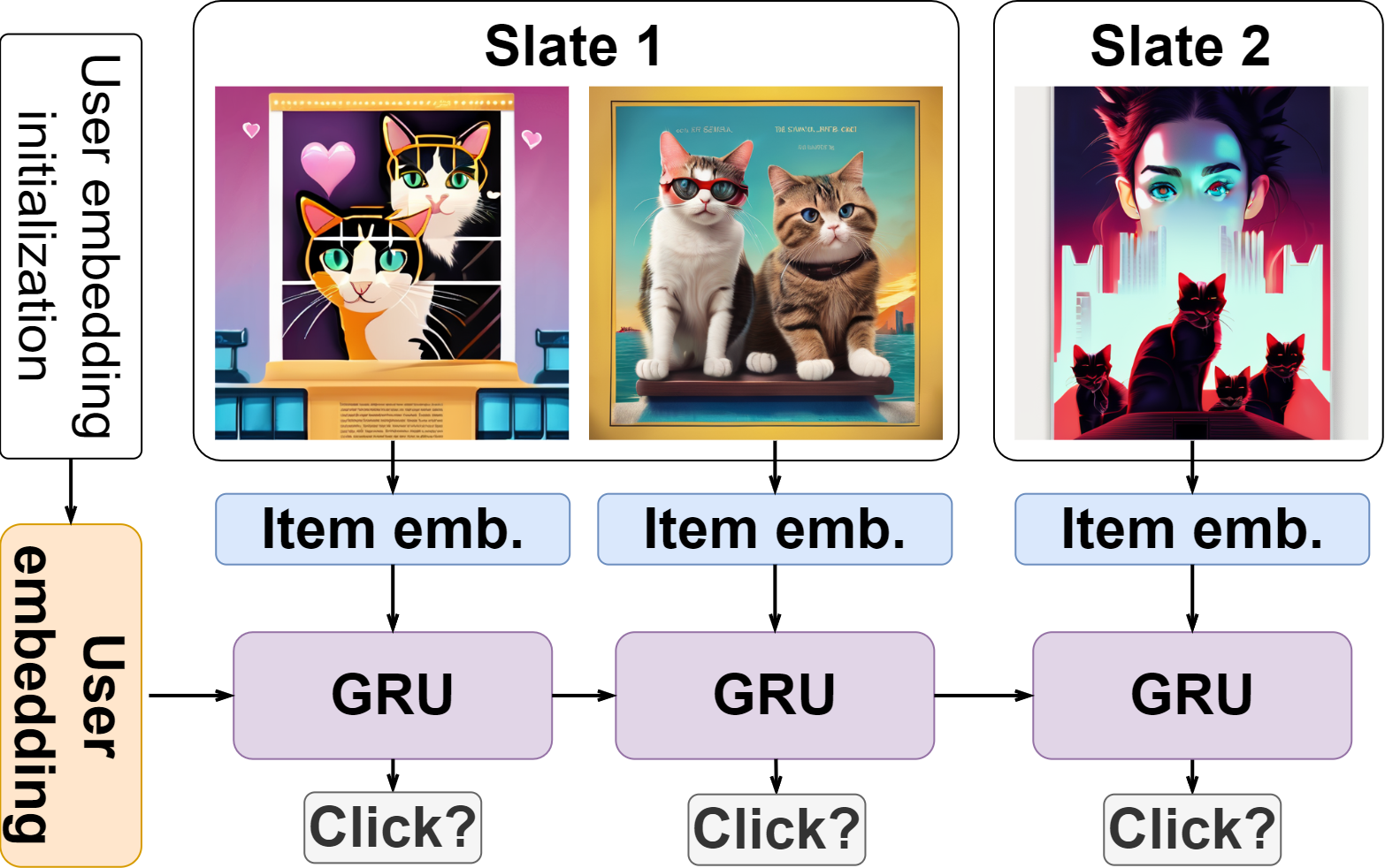} 
&
\includegraphics[height=\figtwohei]{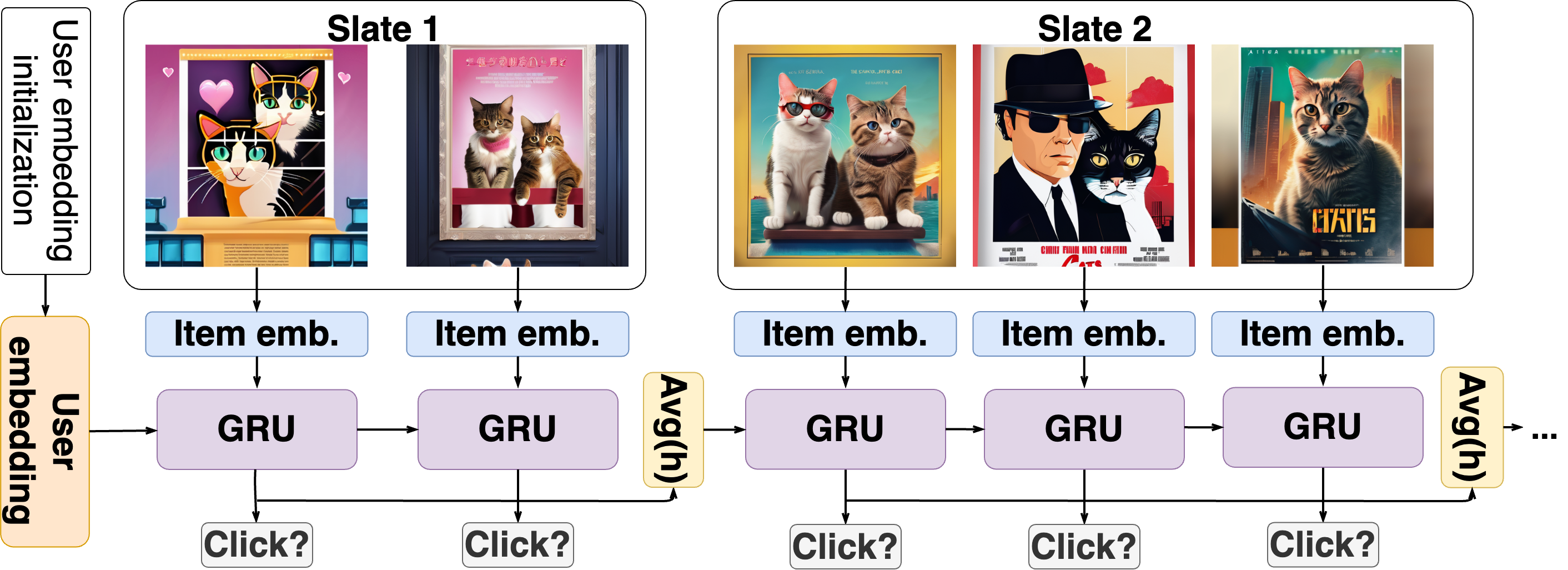} 
\\ (d) Session-wise GRU & (e) Aggregated Slate-wise GRU
\end{tabular}\vspace{.3cm}

\begin{tabular}{c|c}
\includegraphics[height=\figthreehei]{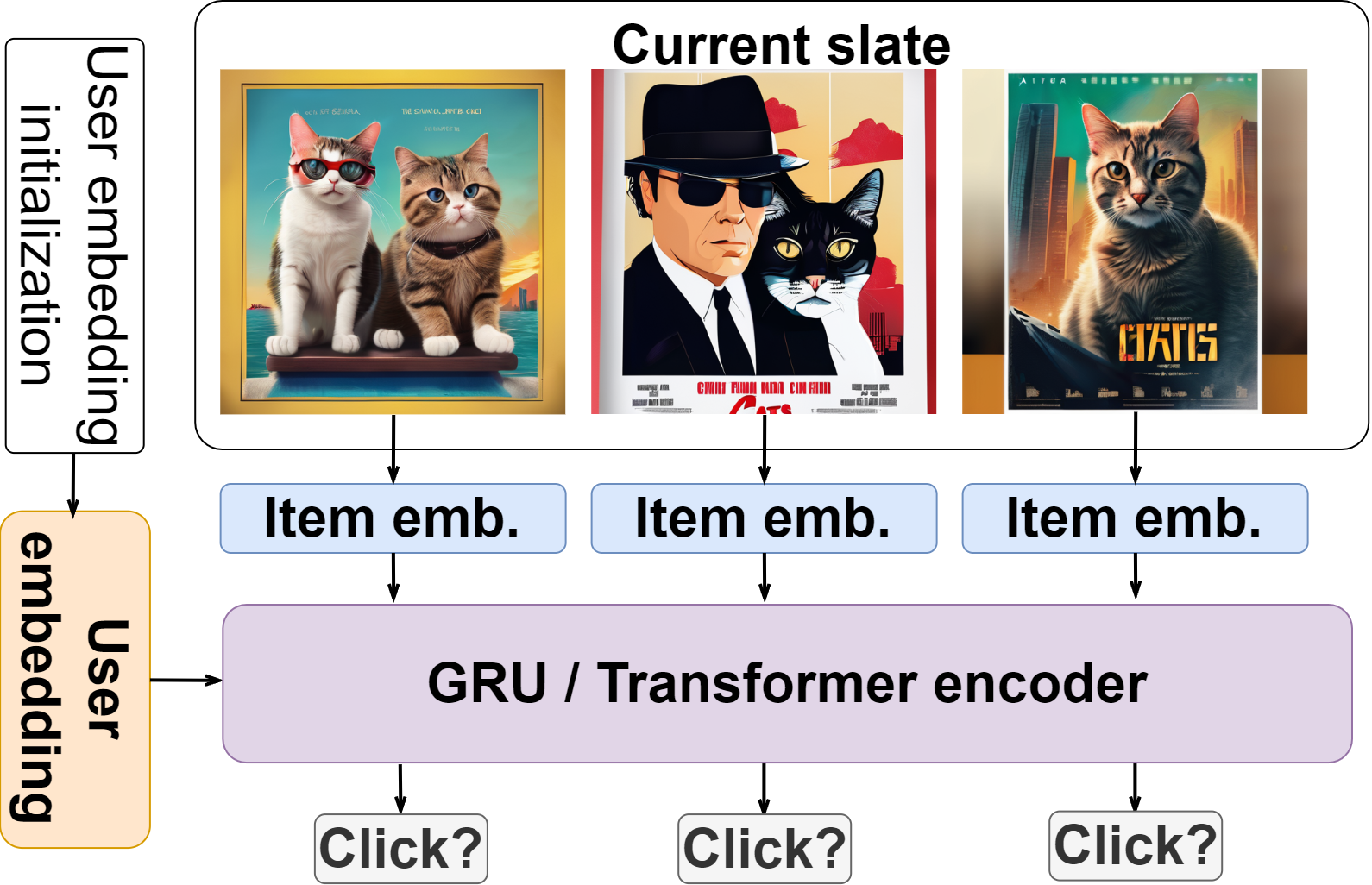} 
&
\includegraphics[height=\figthreehei]{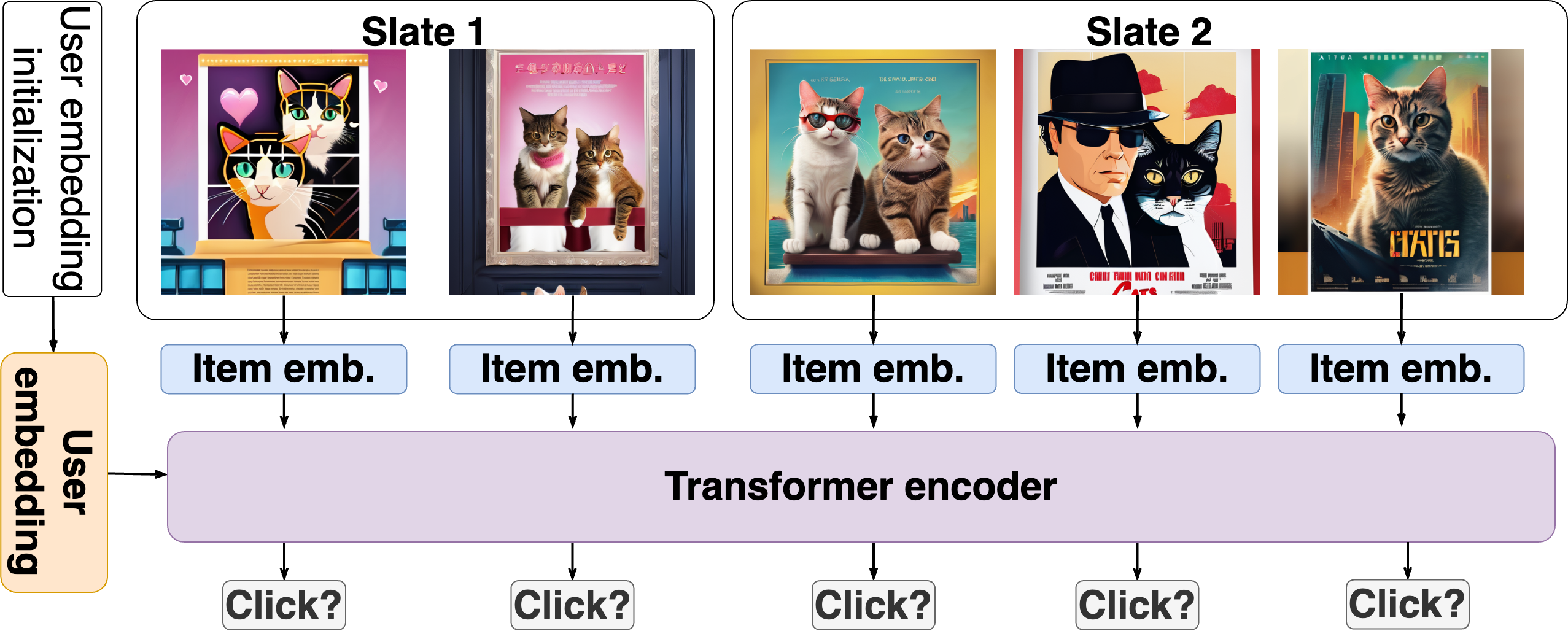} 
\\ (f) Slate-wise GRU / Slate-wise Transformer & (g) Session-wise Transformer
\end{tabular}\vspace{.3cm}

\caption{Datasets and baselines: (a) ContentWise~\cite{contentwise}; (b) RL4RS~\cite{rl4rsdataset}; (c) logistic regression; (d) Session-wise GRU; (e) Slate-wise GRU; (f) Slate-wise GRU/Transformer; (g) Session-wise Transformer. ``Items'' are generated with Stable Diffusion~\cite{Rombach_2022_CVPR}.}\label{fig:db}
\end{figure*}











We have developed baseline models that predict user response with straightforward neural architectures. All models described below can use a pretrained user embedding if one is provided (e.g., in RL4RS);
 if not, we initialize user embeddings with average item embeddings from previous successful interactions of this user (``User embedding initialization'', detailed in Fig.~\ref{fig:db}c). Baselines include:
\begin{inparaenum}[(1)]
\item \emph{matrix factorization},
namely, SVD trained with alternating least squares without reweighing~\cite{KBV09};
\item \emph{logistic regression} 
on concatenated representations (embeddings) of the user and the item (Fig.~\ref{fig:db}c);
\item \emph{Session-wise Gated Recurrent Unit (GRU)},
a recurrent neural network based on a GRU unit~\cite{cho-etal-2014-learning} trained on sessions of individual users (Fig.~\ref{fig:db}d), where user representations initialize the GRU hidden state; this model takes into account the ordering of items shown but ignores the division into slates;
\item \emph{Slate-wise GRU}, trained on individual user sessions 
(Fig.~\ref{fig:db}f), and
\item \emph{Aggregated Slate-wise GRU} that
averages hidden states over a slate before proceeding to the next slate (Fig.~\ref{fig:db}e);
%
%
%
\item \emph{Slate-wise Transformer}, where
a Transformer encoder that receives as input user and item representations and predicts the sequence of clicks (Fig.~\ref{fig:db}f), operating
on every slate independently;
\item \emph{Session-wise Transformer} that reads 
the entire user session rather than just a single slate (Fig.~\ref{fig:db}e); due to the Transformer's quadratic complexity, this model struggles with scaling to real-life (often quite long) sessions.
\end{inparaenum}

\section{Models}\label{sec:models}

Our models developed for user response modeling are in part motivated by click models that simulate user behavior in web search tasks~\cite{chuklin2022click,borisov2016neural}. We note important similarities between the two tasks: users see recommendations arranged as slates, and user responses depend on their previous interaction history and items seen in the current slate. Our models consider either the slate or user session as input and use the user embedding as a condition or hidden state.


The \emph{Neural Click Model} (NCM) is based on a standard recurrent GRU architecture~\cite{cho-etal-2014-learning,hidasi2015session}, just like Slate-wise GRU (Fig.~\ref{fig:db}e), but augmented with the readout technique previously used in NLP~\cite{10.5555/2969033.2969173}, where it means that the token generated at time $t$ is autoregressively fed as input to the RNN at time $t+1$. In NCM, the recurrent unit's output is an embedding vector, and 
 its input is composed of 
    the current item's embedding and
    readout, i.e., information about whether the click has happened at the previous time step, in the form of an item embedding: if there was a click at time $t$ we input the previous item's embedding; if not, the zero vector.

Readout needs to know whether there has been a click on the previous item in the slate. However, we have found that we can improve the results by predicting the entire sequence of clicks at once, sequence-to-sequence. There are several different approaches one could take here:
\begin{inparaenum}[(i)]
    \item choose a threshold for predicted probability $p_i$ to count as a click (positive event);
    \item sample the click event with probability $p_i$;
    \item 
    use the Gumbel-Softmax trick~\cite{jang2017categorical,9729603}, 
    making the entire construction differentiable and suitable for training; we note that apart from reparametrization, this approach also gives an additional source of noise that helps with regularization;
    \item use teacher forcing, substituting clicks from the training set during training; this is a much faster way to train since the gradient does not propagate back beyond a single result;
    however, in our experiments, pure teacher forcing models perform significantly worse.
\end{inparaenum}
In a novel variation for the training process, we propose to use a combination of the Gumbel-Softmax trick with teacher forcing, using full training with readout on some iterations (in our experiments, about $20\%$ were enough) and teacher forcing on others. This significantly speeds up training while preserving model performance.




The \emph{Adversarial Neural Click Model} (AdvNCM)
augments NCM with adversarial training (Fig.~\ref{fig:teaser}a). It has a GRU-based generator and discriminator; during training,
\begin{inparaenum}[(i)]
    \item the generator produces a sequence of clicks for the current sequence of item embeddings;
    \item the discriminator distinguishes it 
    from the real sequence of clicks for the same user and the same item sequence.
\end{inparaenum}
Note that AdvNCM is not a GAN: the generator predicts clicks rather than generates a new object from random noise; the discriminator is used as a source of another loss function.
%
We also apply reparametrization, differentiating the adversarial loss 
via the Gumbel-Sigmoid trick.

\begin{figure}[!t]\centering \includegraphics[width=0.9\linewidth]{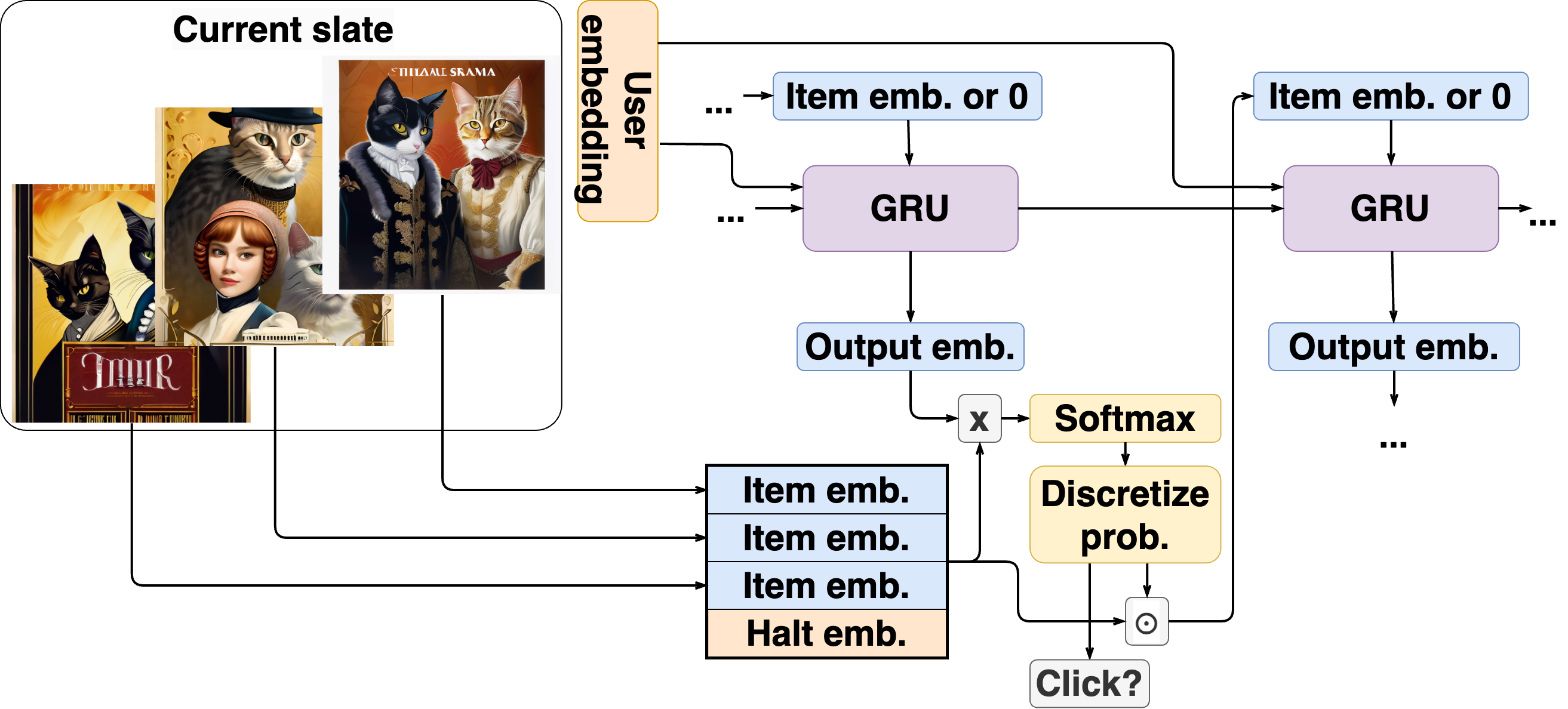}
\caption{The proposed Random Access NCM (RANCM).}\label{fig:rancm}
\end{figure}

\begin{figure}[!t]\centering \includegraphics[width=0.95\linewidth]{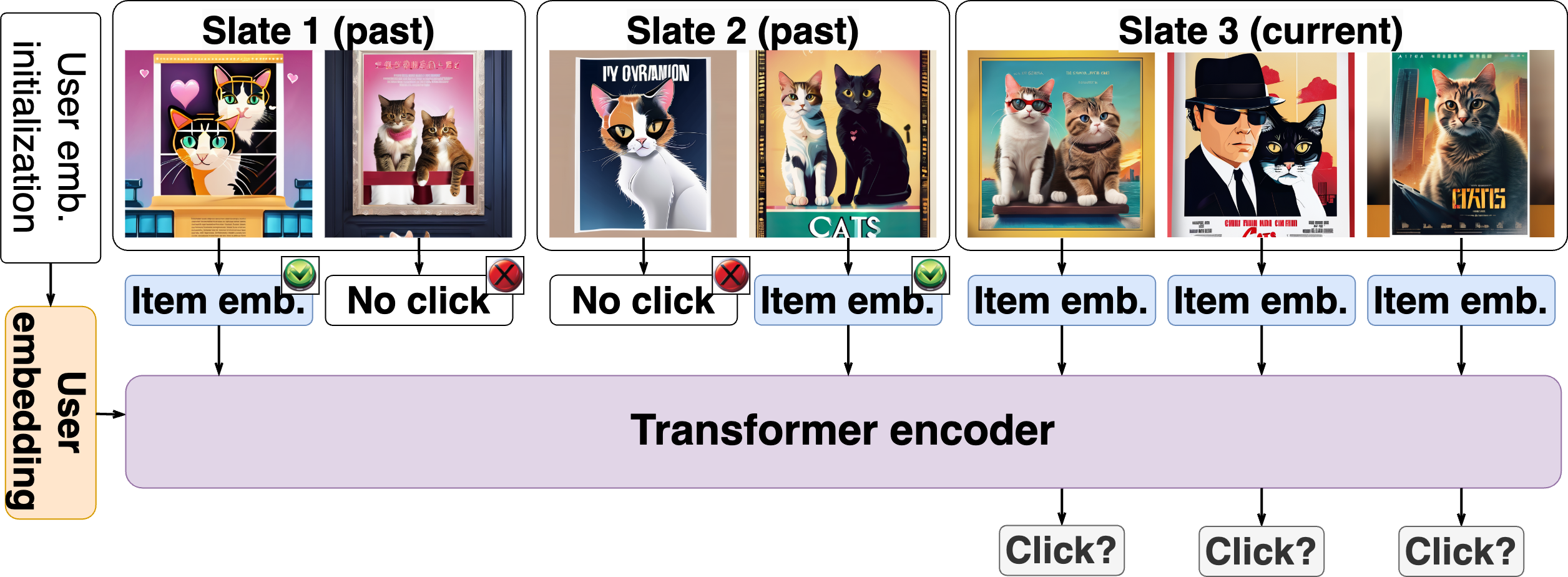}
\caption{The proposed Session-wise Clicked-Only Transformer (SCOT).}\label{fig:trclicked}
\end{figure}

The \emph{Random Access Neural Click Model} (RANCM, Fig.~\ref{fig:rancm}) is inspired by 
classical recurrent models of visual attention~\cite{10.5555/2969033.2969073}; on every step, it
\begin{inparaenum}[(i)]
    \item uses a GRU to produce a vector in the latent space of item embeddings,
    \item multiplies it by the current slate's embeddings with a special learned embedding for the ``Halt'' action, and
    \item processes the resulting scores through the softmax and probability discretization step that can take different forms as discussed for NCM above; if the ``Halt'' action is chosen the process is stopped, otherwise RANCM predicts the next click.
\end{inparaenum}
%
RANCM may be especially convenient as an agent for RL-based experiments.
%
%
\emph{Session-wise Clicked-Only Transformer} (SCOT, Fig.~\ref{fig:trclicked})
is similar to the Session-wise Transformer baseline but inputs only embeddings of items that actually received clicks on previous slates, with
attention masks preserving temporal ordering, so predicting a click event only attends to previous click events. 
This helps reduce complexity since clicked items are a small minority of impressions.




Finally, 
the two-stage \emph{Transformer+GRU} model (Fig.~\ref{fig:teaser}b) combines slate-wise and session-wise models:
\begin{inparaenum}[(i)]
\item a Transformer encoder predicts user clicks on the current slate,
\item its outputs are aggregated, the resulting vector is sent to a GRU unit, and
\item GRU outputs the vector for processing the next slate.
\end{inparaenum}
Here, Transformers are run separately on every slate, without bloating the computational complexity, but at the same time exchanges important information about previous slates in the session via the recurrent mechanism.


\section{Experimental Evaluation}\label{sec:eval}

\begin{table}\centering\setlength{\tabcolsep}{2.5pt}\setstretch{.83}\small
\resizebox{\linewidth}{!}{
\begin{tabular}{p{.2\linewidth}c|rrr|rrr}
\toprule
  \textbf{Model} & \textbf{Emb.} & \multicolumn{3}{c|}{\textbf{ContentWise}} & \multicolumn{3}{c}{\textbf{RL4RS}} \\
& & \multicolumn{1}{c}{AUC} &   \multicolumn{1}{c}{$\mathrm{F}_1$} &  \multicolumn{1}{c|}{Acc.} & \multicolumn{1}{c}{AUC} &   \multicolumn{1}{c}{$\mathrm{F}_1$} &  \multicolumn{1}{c}{Acc.} \\
\midrule
MF &  & 0.654 & 0.198 & 0.257 & 0.719 & 0.755 & 0.653 \\

Logistic & SVD & 0.625 & 0.246 & 0.784 & 0.863 & 0.846 & 0.790 \\
regression & NN & 0.689 & 0.270 & 0.787 & 0.915 & 0.880 & 0.837 \\
 & Ext. & --- & --- & --- & 0.863 & 0.844 & 0.769 \\
 

Slate-wise & SVD & 0.669 & 0.274 & 0.887 & 0.900 & 0.868 & 0.818 \\
Transformer& NN &  0.740 & 0.328 & 0.839 & 0.930 & 0.890 & 0.853 \\
& Ext. & --- & --- & --- & 0.925 & 0.888 & 0.848 \\
 
Session-wise & SVD & 0.637 & 0.261 & 0.817 & 0.881 & 0.849 & 0.782 \\
Transformer & NN & 0.704 & 0.284 & 0.774 & 0.925 & 0.885 & 0.843 \\
 & Ext. & --- & --- & --- & 0.914 & 0.876 & 0.831 \\



Slate-wise & SVD & 0.680 & 0.280 & 0.804 & 0.901 & 0.867 & 0.826 \\
GRU & NN & 0.744 & 0.331 & 0.860 & 0.924 & 0.878 & 0.832 \\
 & Ext. & --- & --- & --- & 0.923 & 0.884 & 0.845 \\
 
Aggregated & SVD & 0.657 & 0.230 & 0.870 & 0.860 & 0.851 & 0.796 \\
 Slate-wise & NN & 0.725 & 0.300 & 0.826 & 0.851 & 0.850 & 0.781 \\
 GRU & Ext. & --- & --- & --- & 0.856 & 0.850 & 0.789 \\


 Session-wise & SVD & 0.678 & 0.261 & 0.859 & 0.863 & 0.844 & 0.793 \\
 GRU & NN & 0.740 & 0.308 & 0.739 & 0.859 & 0.851 & 0.790 \\
 & Ext. & --- & --- & --- & 0.861 & 0.847 & 0.793 \\
 

\midrule
Neural & SVD & 0.672 & 0.257 & 0.890 & 0.908 & 0.868 & 0.835 \\
Click Model& NN & 0.730 & 0.283 & \bfseries 0.891 & 0.919 & 0.884 & 0.848 \\
 & Ext. & --- & --- & --- & 0.916 & 0.882 & 0.848 \\

AdvNCM & SVD  & 0.719   & 0.348     & 0.860     & 0.905 &  0.868 & 0.816 \\
       & NN & \bfseries 0.791 & 0.351 & 0.858 & 0.906 &  0.868 & 0.820 \\
       & Ext. & --- & ---   & ---   & 0.869 & 0.837 & 0.762 \\

RANCM  & SVD  & --- & \bfseries 0.362 & 0.871     & --- & ---   & ---   \\
       & NN   & --- & 0.346 & 0.860 & --- & ---   & ---   \\


SCOT & SVD & 0.672 & 0.297 & 0.864 & 0.910 & 0.875 & 0.831 \\
 & NN & 0.715 & 0.306 & 0.813 & \bfseries 0.934 &  \bfseries 0.896 & \bfseries 0.865 \\
 & Ext. & --- & --- & --- & 0.927 & 0.891 & 0.860 \\
 
Transformer & SVD & 0.625 & 0.193 & 0.248 & 0.859 & 0.842 & 0.791 \\
+ GRU & NN & 0.737 & 0.317 & 0.850 & 0.877 & 0.856 & 0.808 \\
 & Ext. & --- & --- & --- & 0.872 & 0.854 & 0.804 \\
\bottomrule
\end{tabular}}

\caption{Experimental results (best in bold);
SVD~-- MF embeddings; NN~-- learnable emb. layer; Ext.~-- external embeddings.}\label{tbl:res}\vspace{-.5cm}
\end{table}

Table~\ref{tbl:res} summarizes the main experimental results on the \emph{ContentWise} and RL4RS datasets. We compare the baselines and newly developed models across three different metrics: ROC-AUC, $F_1$ score, and accuracy (only RL4RS provides external embeddings, but it does not have click order so RANCM does not work there). Throughout Table~\ref{tbl:res}, the leaders in ROC-AUC and $F_1$ score have statistically significant differences compared to other models, while accuracy is a much noisier metric, so our conclusions are mostly based on AUC and $F_1$.
%
The proposed models consistently outperform natural baselines on both datasets, with the best baseline being Slate-wise Transformer. In particular,
    on \emph{ContentWise} the best model is AdvNCM with RANCM close behind in terms of $\mathrm{F}_1$, 
    but on RL4RS the situation is different, with SCOT being best. This is possibly due to the high fraction of positive interactions in RL4RS; this is definitely the reason for low RANCM performance: on RL4RS it usually halts far too early.
    Second, on average trainable embeddings improve the results. 
    Third, there is a significant gap between GRU baselines and NCM which is based on the same architecture; this confirms our hypothesis that readout is an important improvement.
%

\section{Conclusion}\label{sec:concl}

In this work, we have proposed neural architectures for user feedback modeling based on click models,
significantly improving the results over baselines on two standard datasets, \emph{ContentWise} and RL4RS. We note that RNN-based models are quite competitive with Transformer-based ones, and the best architecture we found is an adversarial recurrent network. Results, however, heavily depend on the dataset and evaluation metric. Learnable neural embeddings improve metrics on average, but usually only slightly, so
in practical applications SVD-based embeddings are still a reasonable first step. In future work, our models can extend the range of user actions and can provide environments for offline RL approaches.



\begin{acks}
This work was supported by a grant for research centers in the field of artificial intelligence, provided by the Analytical Center for the Government of the Russian Federation in accordance with the subsidy agreement (agreement identifier 000000D730321P5Q0002) and the agreement with the Ivannikov Institute for System Programming of the Russian Academy of Sciences dated November 2, 2021 No. 70-2021-00142.
\end{acks}

\bibliographystyle{ACM-Reference-Format}
\bibliography{sample-base}

\end{document}